# Gibbs adsorption impact on a nanodroplet shape:
## modification of Young-Lapplace equation

Mykola Isaiev[1,2], Sergii Burian[2], Leonid Bulavin[2], William Chaze[1], Michel Gradeck[1], Guillaume Castanet[1], Samy Merabia[3], Pawel Keblinski[4], Konstantinos Termentzidis[1]

[1]LEMTA, CNRS-UMR7563, Université de Lorraine, Vandoeuvre les Nancy, F-54500, France

[2]Faculty of Physics, Taras Shevchenko National University of Kyiv, 64/13, Volodymyrska Str., Kyiv, Ukraine, 01601

[3]Université de Lyon 1, ILM, CNRS-UMR5306, 69621 Villeurbanne, France

[4]Department of Materials Science and Engineering, Rensselaer Polytechnic Institute, Troy, NY 12180, United States

E-mail: mykola.isaiev@knu.ua
E-mail: konstantinos.termentzidis@univ-lorraine.fr

**Keywords:** nanoscale droplet, external field, molecular dynamics, adsorption, Young-Laplace equation

An efficient technique for the evaluation of the Gibbs adsorption of a liquid on a solid substrate is presented. The behavior of a water nanodroplet on a silicon surface is simulated with molecular dynamics. An external field with varying strength is applied on the system to tune the solid-liquid interfacial contact area. A linear dependence of droplet's volume on the contact area is observed. Our modified Young-Laplace equation is used to explain the influence of the Gibbs adsorption on the nanodroplet volume contraction. Fitting of the molecular dynamics results with these of an analytical approach allows us to evaluate the number of atoms per unit area adsorbed on the substrate, which quantifies the Gibbs adsorption. Thus, a threshold of a droplet size is obtained, for which the impact of the adsorption is crucial. Moreover, the presented results can be applied for the evaluation of the adsorption impact on the physical-chemical properties of systems with important surface-to-volume fraction.

## 1. Introduction

The interfaces between two or more different phases predominate a variety of physical and chemical properties, especially at the nanometer scale compared to bulk materials, as the surface-to-volume fraction become very high. In particular, interactions between different phases are crucial for the wettability, solubility, chemical reactions etc. The prediction of the interaction of different phases is a key challenge for a wide range of applications in chemical industry, catalysis, material science, microfluidics, pharmacology, medicine, etc.

As an example, we can note, the wetting properties of a liquid droplet on a solid substrate. For the case of nanoscale droplet additionally to the surface tension, the line tension influence on the wetting angle should be taken into account. These excesses of free energy per unit length separating three distinct phases respectively should be taken into account to achieve stability of deformable surfaces. The general Young equation can be presented as follows:

$$\cos(\theta_w) = \frac{\gamma_{vs} - \gamma_{ls}}{\gamma_{vl}} - \frac{\tau}{\gamma_{vl}} \frac{1}{r},\tag{1}$$

where $\theta_w$ is the wetting angle, $\gamma_{vs}, \gamma_{ls}$, and $\gamma_{vl}$ are the vapor/solid, liquid/solid, and vapor/liquid surface tensions respectively, $\tau$ is the line tension, and $r$ is the contact radius of a three-phase line. Barisic and Beskok [1] have simulated spherical droplets with different volumes located on the silicon substrate with molecular dynamics. They measured a wetting angle as a function of the curvature of three phase contact line ($1/r$). With the use of a linear dependence (**Equation** (1)), they estimated the line tension and they defined several parameters of the interaction between water molecules and silicon atoms with the use of the experimental wetting angles of macrosize droplet as input. Later on, it was shown in [2] that the simulation of a cylindrical droplet is more efficient than this of spherical droplets for the evaluation of the interaction parameters between a droplet and a substrate, since the curvature of the three-phase contact line is equal to zero [3] in the case of cylindrical droplets. Leroy et al[4] have shown recently that the evaluation of the force-fields parameterization using experimental wetting angles of droplet has limitations. They have stated that the dependence of water surface tension on the chosen water model should be taken into account. Thus, additional markers for reliable description of water-solid interfacial properties should be found, such as "work of adhension"[4]. Numerous MD studies [5–8] of solid-liquid interfaces refer to the density fluctuations of fluids near the solid free surfaces. These fluctuations appear as a result of molecules or atoms absorbed in or adsorbed on the solid substrate. The density of adsorbents strongly depends on how these atoms or molecules interact with the substrate. The density variation can even lead to experimentally perceptible contraction of the volume of the liquid containing nanoparticles [9, 10]. Moreover, the altered structure of confined water near the wall can lead to a drastic change of its physical properties [11–13].

An important question that one should address in all cases, is if one can treat nanoscale droplets with macroscopic approaches? Which are the limitations of such approaches, or what should be the corrections to achieve convergence of dimensions? In this paper we focus on the influence of the absorbed layer on the droplet shape and droplet volume. We study the contraction of a nanoscale cylindrical water droplet on a silicon substrate with means of MD. We applied a homogeneous external field with different intensities to tune the contact area. We correlate the shape of a nanoscale droplet modeled and simulated by MD with the one obtained by a macroscopic method of hydrostatic similarity of droplet shape in gravity field. The deviations between the results of the MD simulations and the analytical approach will be explained. Furthermore, we propose a new model accounting the contraction of the droplet volume due to adsorption phenomena. The proposed model can be also relevant of experimentally observed volume contraction of nanocolloidal solutions [9, 10] or nanoporous materials filled with liquids [14, 15]. These systems are accessible to experimental measurements and they have high ratio of surface/volume.

## 2. Molecular dynamics simulations of water droplet on silicon substrate

### 2.1. MD Methodology

All MD simulations presented here are performed using LAMMPS code (Large-scale Atomic/Molecular Massively Parallel Simulator) [16]. We have considered cylindrical droplet on a smooth silicon substrate with and without an external homogeneous field. Additionally, cylindrical droplet without substrate was simulated as reference. We use cylindrical droplets to avoid the influence of line tension on the contact angle. In both cases the simulation box dimensions are $108.6 \times 217.2 \times 200$ Å in x, y and z-direction respectively.

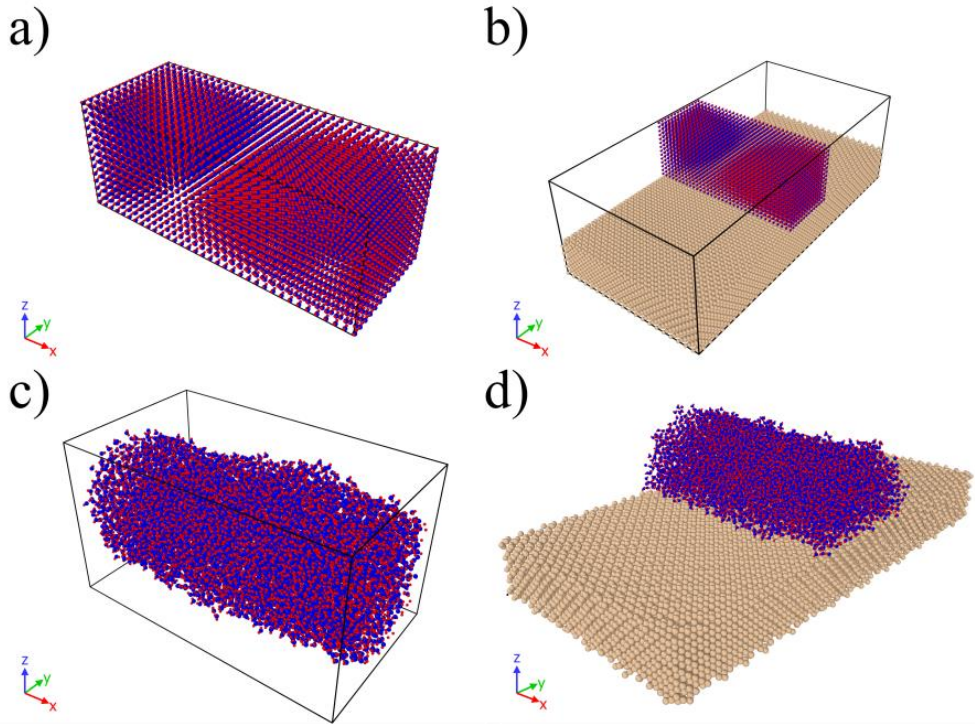

**Figure 1.** Initial (a,b) and after thermostating (c,d) configurations of a cylindrical nanoscale water droplet for the cases of a droplet (a,c) without substrate and a droplet on a silicon substrate (b and d).

The initial configurations of the atoms of the droplet without/with substrate are presented in **Figure 1a** and **1b** respectively. Initially, a rectangular crystalline water droplet with size $108.6 \times 43.4 \times 43.4$ Å was built. The total number of water molecules in the droplet is equal to 6860. The distance between oxygen atoms of water molecules was set to 3.1 Å, and this represents approximately the density of the bulk water under normal conditions. The surface of the crystalline silicon substrate has the orientation of the (1, 0, 0) facet, with lattice constant equals to $a_0 = 5.43$ Å. The silicon slab has dimensions $108.6 \times 217.2 \times 27.15$ Å ($40 \times 80 \times 10$ atoms).

Interactions between water molecules are described by the extended simple point charge (SPC/E) water model [17], with parameters taken by Orsi [18]. The 12-6 Lenard-Jones interatomic potential was used for the interactions between oxygen and silicon atoms. The parameters of the interaction were chosen as $\varepsilon_{Si-O} = 15.75$ meV and $\sigma_{Si-O} = 2.6305$ Å [2]. In our study we have neglected the interactions between hydrogen and silicon atoms. The interactions between silicon atoms the Stilling-Weber potential [19] is used. Additionally, the harmonic force was applied to atoms of the lower four atomic layers ($z < 2a_0$) to tether them to their initial positions. This is done to prevent moving the silicon slab in Z direction during the simulations.

The initial configurations were thermostated during 2 ns with the Nosé–Hoover thermostat in the canonical ensemble to achieve equilibration at 300 K. The timestep was set equal to 1 fs. The resulting droplets are presented in **Figure 1c** and **1d** for the free droplet and the droplet on the substrate.

### 2.2. Droplet without external field

The water density profiles were collected every 0.1 ps and then averaged during 0.5 ns. To obtain the density profile, the three-dimensional space was divided by square cross-section rectangular ($1 \times 1$ Å$^2$, in YZ plane). The resulting density profile of the free droplet and the dependence of the density as a function of the distance from the droplet centre are presented in **Figure 2a** and **2b** respectively.

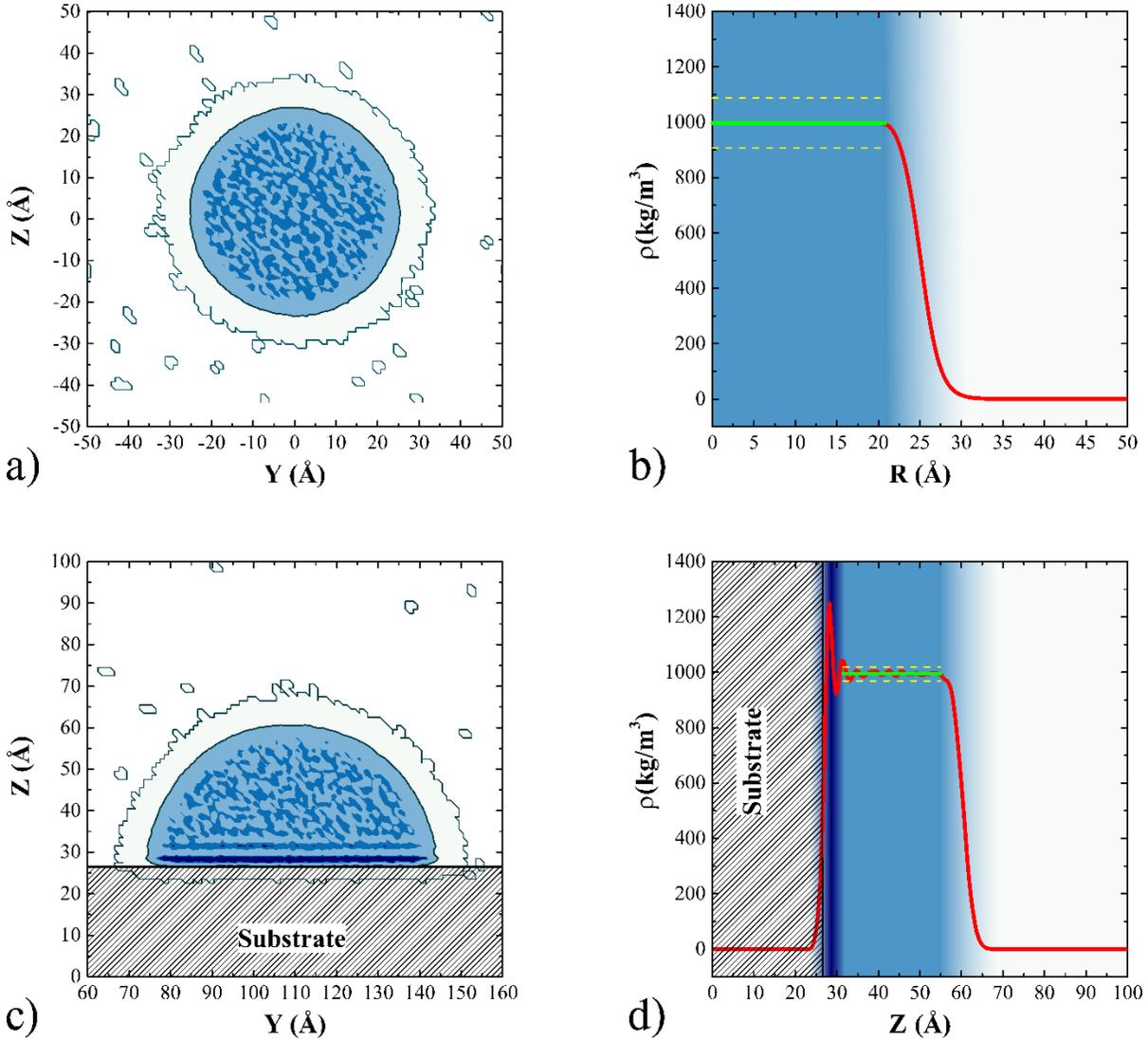

**Figure 2.** a) Averaged density map of a free droplet; b) Droplet density profile as a function of the distance from the droplet centre; c) Averaged density map of a droplet on a silicon substrate; d) Averaged density profile of a droplet in the plane perpendicular to y-axis and crossing the droplet's centre of mass as a function of z.

The density of bulk water inside the droplet was found to be equal to 997.6 kg/m$^3$ and the density of the equilibrium vapor is 0.01474 kg/m$^3$ and pressure equal to 0.05 atm. These densities are in a good agreement with the results published previously for the same water model [20] and with experimental data. The liquid-vapor interface of the droplet is relative sharp (**Figure 2a** and **2b**), and it has thickness approximately equal to 10 Å. We define the liquid/vapor interface as a constant density surface equal to the half of the bulk water density. In **Figure 2a** and **2b** the liquid/vapor interfaces are presented as bold black solid lines. The radius of the droplet is approximately equal to 25.1 Å and its volume per unit length is $v = 1977.6$ Å$^2$ (**Figure 2a** and **2b**). Additionally, in **Figure 2c** the density map of the droplet on the silicon substrate in the case of absent of external forces is depicted. We have evaluated the wetting angle which has been found to be equal to 86.6$^\circ$. This value is in a good agreement with previous results [2]. In **Figure 2d** the dependence of the averaged water density in the plane perpendicular to Y-axis and crossing the droplet's centre of mass as a function of z is presented. As it was mentioned above, a layering effect appears close to the substrate. The density of this layer can be higher than the density of bulk water under normal conditions for more than 30 %. The thickness of this layer is calculated to be approximately 4±1 Å. The presence of this layer leads to the droplet volume contraction, as the volume per unit length in this case becomes equal to 1803.1Å$^2$ (without the presence of the substrate it was found to be equal to 1977.6 Å$^2$)

### 2.3. The impact of the external homogeneous field

In order to study the interaction between a liquid and a solid in an external homogeneous field, we added gravity-like field, with which all water atoms are subject to an additional force in direction towards the negative values of the z-axis. This additional field was added during the thermalization process described above. In **Figure 3a** and **3b** MD snapshots of the droplet in the external field with intensities $10.1\times10^{12}$ and $51.3\times10^{12}$ m/s$^2$ are depicted. These values correspond to radius of curvature 2 and 64 times greater than for the case of absence of gravity. We are aware that these gravity fields are enormous, but we can justify them by two reasons. First of all, as it was explained in the introduction our aim is to reproduce macroscopic behavior of droplets with means of molecular dynamics.

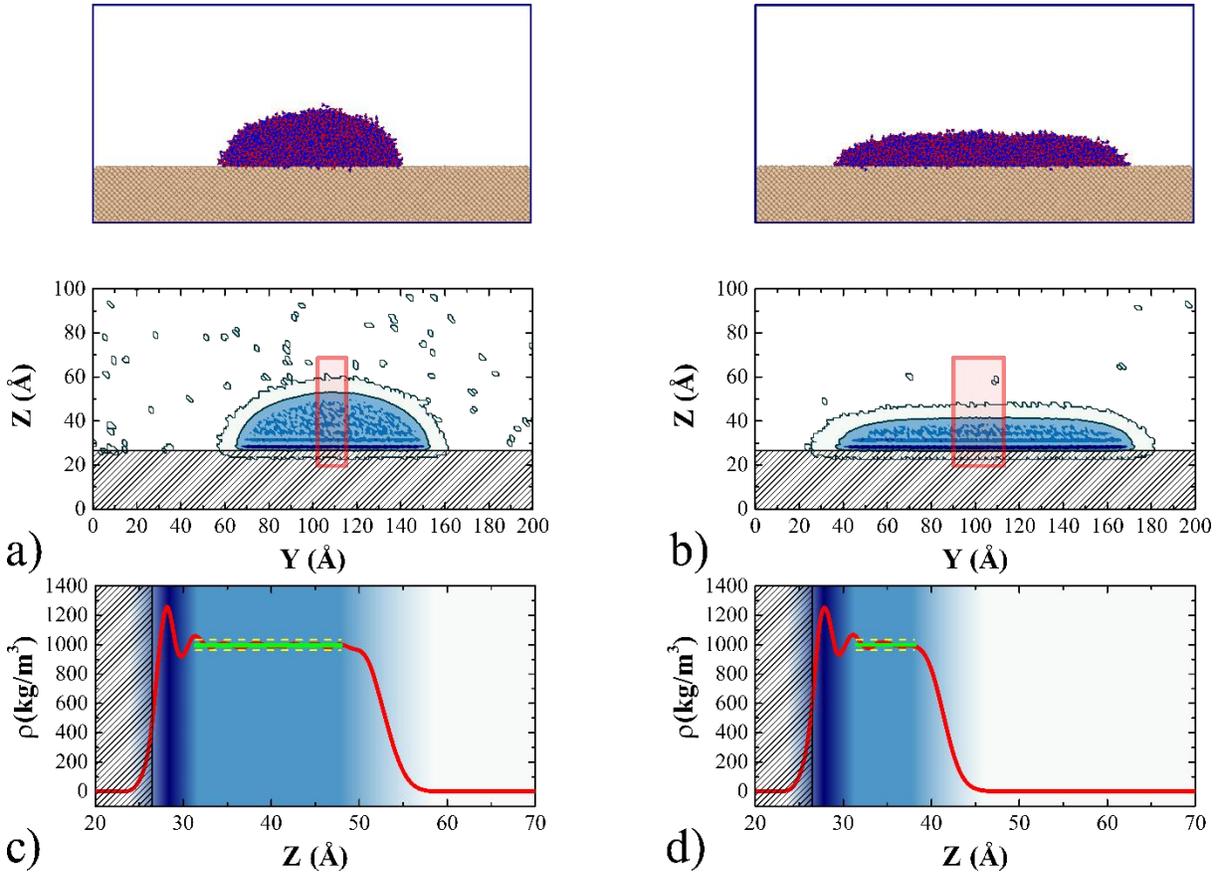

**Figure 3.** Molecular dynamics snapshots and the density profiles (upper insets) of the droplet density map in an external field with intensities $10.1 \times 10^{12}$ (a) and $51.3 \times 10^{12}$ N/kg (b), corresponding density profiles in the middle of water droplet for the same intensities of external field (c) (d)

This might be useful for several studies involving two or three phases interfaces and their dynamic interactions. The second reason is related to AFM or SThM measurements. The tip of an AFM can exercise huge forces on a water meniscus (formed around a tip) or a solid surface. These forces could be of the order of F = 100 pN. If we divide this force with the mass of a water molecule, we obtain acceleration of the order of $10^{13}$ m/s$^2$, which is in the same order of the imposed external field.

In the upper insets of **Figure 3a** and **3b**, the averaged density profiles of the droplets under the influence of the external field are presented. It is important to note that the thickness of the adsorbed layer and its density is approximately the same for all cases and equal to the one without external field. The dependence of the contact area as a function of the external field magnitude is plotted in **Figure 4a**.

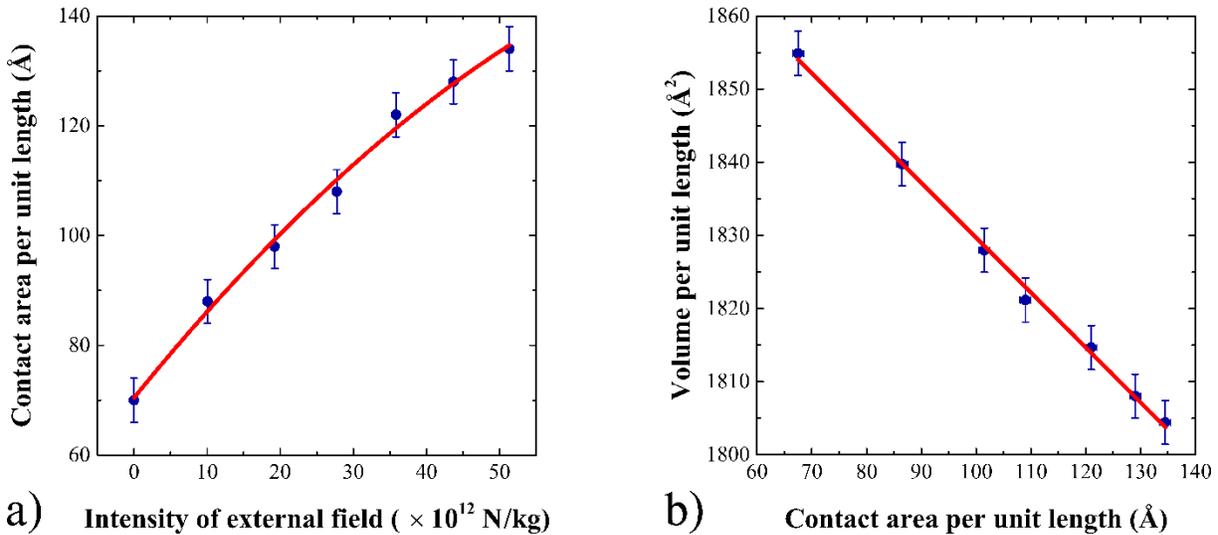

**Figure 4.** a) The dependence of the contact length on the external field intensity; b) The dependence of the droplet volume per unit length of cylinder as a function of the contact area per unit length

The contact area increases by 50 % in adding an external field with magnitude 50 $10^{12}$ m/s$^2$. Due to the increase of the contact area the fraction of the water molecules, which are in contact with the substrate, increases. Consequently, the number of the water molecules located inside the force field of the silicon atoms is also increased. Since this field leads to the density increase of the adsorbed atoms, the total volume of the droplet decreases. The dependence of the droplet volume on the contact area is presented in **Figure 4b**. The dependence is linear, which proves that the change of the water volume arises due to the increase of the number of water molecules adsorbed on the silicon surface. The fitted equation for the dependence can be presented as follows:

$$\tilde{v} = v - \varepsilon a \,, \tag{2}$$

where $\tilde{v}$ is a volume of contracted droplet, $a$ is the contact area per unit length, $\varepsilon$ is the slope of the dependence of volume on the contact area $\tilde{v}(a)$(**Figure 4b**). We fitted the results of the molecular dynamic simulations with the equation (2), and found that the best fitting value is $\varepsilon = 0.875$ Å.

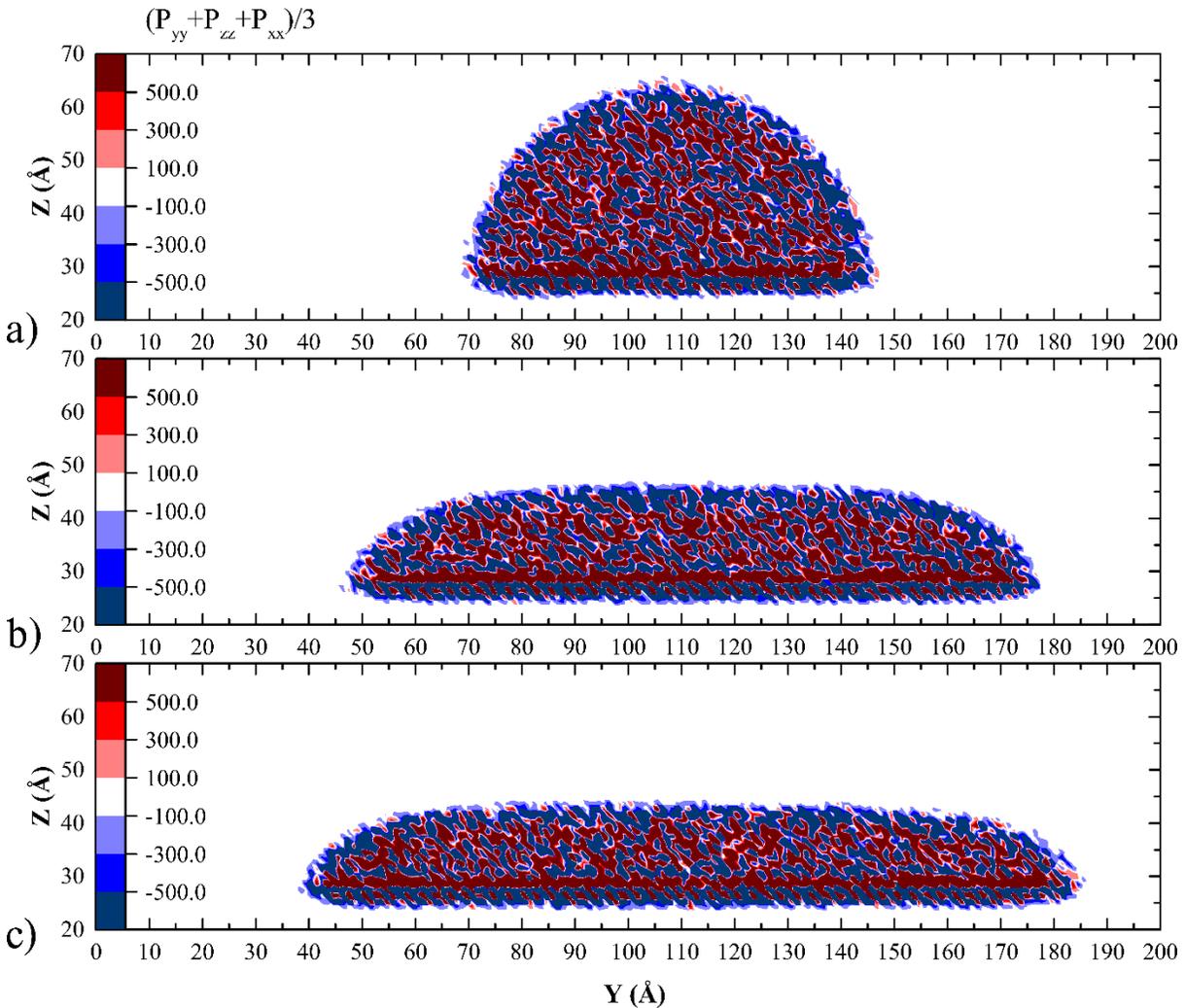

**Figure 5.** Pressure maps of the droplet under absence (a) and presence of external force field 20 (b) and 50 (c) $10^{12}$ N/kg. No gradient of pressure is observed in any case.

From the estimated ε, the surface excess or adsorption per unit area Γ [21] can be calculated (in mol/m$^2$) as follows

$$\Gamma = \frac{\varepsilon \rho_b}{M_{H2O}},$$ (3)

where ε should be taken in m. Then, we estimate that Γ equals to 4.8 μmol/m$^2$.

At this point, it is important to check possible effects due to capillary forces. If capillary forces were the origin of the phenomena observed in our case, one could have observed a gradient of water pressure in the z direction. This is not the case as it can been seen if **Figure 5**, where one can observe that indeed the pressure is higher close to the water/silicon interface (this is related to the adsorption layer), but in the main volume of the water droplet we observe that there is no systematic way that the pressure varies along the z-direction.

### 3. Analytical model of a nanodroplet in an external field

In the previous section the simulation of the nanoscale droplet in an external field with MD is presented. In this section an analytical approach of the droplet volume contraction due to the enhancement of the contact area will be presented. As it was mentioned previously, we assume that the volume of a droplet decreases due to the influence of the potential field of the atoms of the substrate. Direct approach requires calculation of the field created by the atoms of the substrate. Since this field is strongly heterogeneous, the calculation of the droplet shape can be very cumbersome from practical point of view. Therefore, we consider the influence of this field phenomenologically due to considered adsorption of water molecules on silicon surface.

For the quantitative description of the droplet shape in an external field at the macroscale the approach based on the Young-Laplace equation is usually used [22, 23]. The Young-Laplace equation describes the mechanical equilibrium condition for two homogeneous fluids separated by an interface that is assumed to be a *surface* of *zero thickness*. In the case of the absence of external field, the pressures difference $\Delta P$ across a curved interface caused by the surface tension can be presented as follows

$$\gamma \left( \frac{1}{R_1} + \frac{1}{R_2} \right) = \Delta P, \tag{4}$$

where, $\gamma$ is the surface tension, $R_1$ and $R_2$ are the principal radii of the surface curvature. For the considered case of the cylindrical droplet one of the principal curvature radius is trend to be equal infinity ($R_2 \to \infty$).

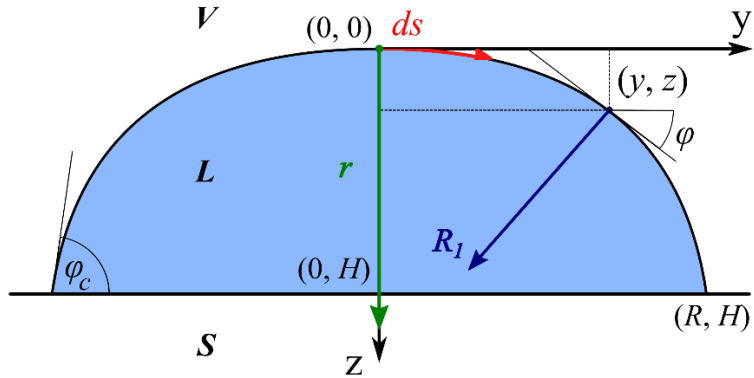

**Figure 6.** A schematic representation of the geometry of the studied droplet

In the frame of the assumption of the incompressible isotropic fluid in an external homogeneous gravity-like field, the equation of the force equilibrium can be presented as follows[24]

$$\Delta P = \Delta P_0 + \Delta \rho g (z - z_0), \tag{5}$$

where $\Delta P_0$ is the pressure difference in reference point $z = z_0$, $\Delta \rho$ is the difference in densities between the two phases, and $g$ is the external field intensity. In the Equation 5 the direction of the z-axis is chosen the same as it is presented in **Figure 6** (invers to the direction of z -axis in MD simulation).

In the case when the origin of the coordinate system and the reference point are located at the droplet apex, the equation (5) can be presented as follows

$$\frac{1}{R_1} = \frac{1}{r} + \frac{z}{\kappa^2},$$ (6)

where r is the radius of curvature at the droplet apex, $\kappa = \sqrt{\gamma/(\Delta \rho g)}$ is the capillary length. In this case we neglected the influence of the size dependence of the surface tension [25].

### 3.1. Reflectional symmetric drop shape analysis

For the description of a droplet shape in a gravity field, Axisymmetric Drop Shape Analysis (ADSA)[26] proposed by Rotenberg et al is most commonly used. Regarding cylindrical droplet, the modification of ADSA can be used taking into account reflectional symmetry of the droplet shape for the analysis (RDSA). In this case the curvature can be related to the arc length s and the angle of inclination of the interface to the horizontal angle $\varphi$ by:

$$\frac{1}{R_1} = \frac{d\varphi}{ds}.$$ (7)

Therefore, the equation (4) can be presented as follows

$$\frac{d\varphi}{ds} = \frac{1}{r} + \frac{z}{\kappa^2}.$$ (8)

The equation (6) is supplemented with the geometrical relations

$$\frac{dy}{ds} = \cos(\varphi); \frac{dz}{ds} = \sin(\varphi); \frac{dv}{ds} = 2y\sin(\varphi),$$ (9)

where $v$ is the droplet volume per unit length. The system formed by Equation 7, 8 and 9 is solved numerically using the Runge-Kutta method taken as initial conditions $y(0) = z(0) = \varphi(0) = v(0) = 0$. Integration is stopped when $\varphi$ reaches the value of the contact angle $\varphi_c = 86.6^o$ that is a priori known parameter.

The dependence of the droplet height and its contact radius as a function of the gravity intensities evaluated numerically (filled squares) and with MD simulation (filled circles) are presented in Figure 7a and 7b respectively. As one can see, there is a deviation between results of the numerical approach and results of MD simulations. This discrepancy might arise because of the presence of the adsorbed water layer.

Since the physical basis of the Equation 4 is the assumption of the incompressibility of the liquid, the volume of a droplet under different gravities is the same. Although the volume of the macrosize droplet conserved in the presence of gravity field, the volume of the nanoscale droplet is change due to the presence of the adsorbed layer. To take this into account we decomposed the droplet into two parts: adsorption layer and the droplet with capillary shape. In this case the mass of the droplet $m_d$ per unit of length can be determined by:

$$m_d = \rho_b(v - v_a) + \rho_a v_a, \tag{10}$$

where $\rho_b$ and $\rho_a$ are the densities of the bulk water and adsorbed layer respectively; $v_a$ is the volume per unit length of the adsorbed layer. To allow fair comparisons with the MD simulations, the mass of the droplet was kept the same. For that, the parameter $r$ in Equation 8 was varied iteratively. The value of $r$ is increased if $m_d$ is bigger than the requested liquid mass, or decreased in the contrary.

The result of the numerical simulations of the droplet contact radius and radius at the droplet top considering the adsorbed layer are presented in **Figure 7a** and **7b** with green squares.

The values of the adsorbed layer thickness ($H_a = 4$ Å) and its density (1220 kg/m$^3$) was taken from the results of MD simulations.

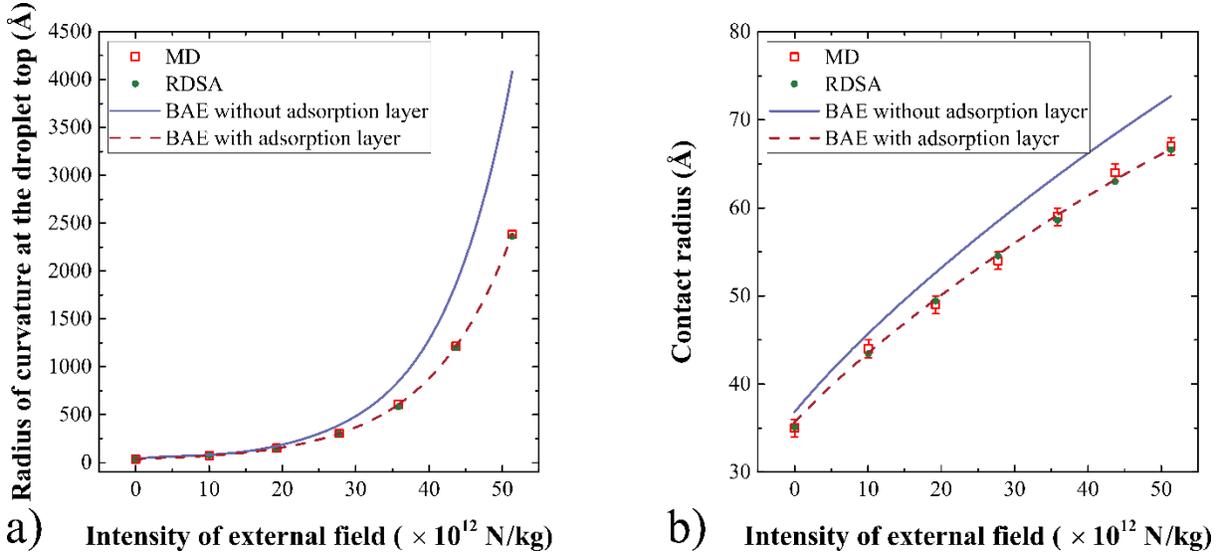

**Figure 7.** Dependences of the droplet height (a) and contact radius (b) as a function of the intensity of gravity field

As one can see in figure-7, there is an excellent agreement of the results of numerical integration of Eq (7) and (8) with the results of molecular dynamics simulations. Therefore, a significant impact of the adsorbed layer presence on the droplet shape contraction can be stated. The Gibbs adsorption in this case can be estimated as follows

$$\Gamma = \frac{H_a(\rho_a - \rho_b)}{M_{H2O}} = 4.9 \ \frac{\mu mol}{m^2}, \tag{11}$$

where $M_{H2O}$ is the molar mass of water. The $\Gamma$ values predicted with Equation 3 and 11 are in excellent agreement.

### 3.2. Bashforth-Adams equation with adsorption layer

Another approach for the droplet shape treatment is based on the Bashforth-Adams equation (BAE) solution. This equation can be obtained from the Equation 6, after replacement of the curvature with expression in the following form:

$$\frac{1}{R_1} = \frac{d^2z}{dy^2} \Big/ \left(1 + \left(\frac{dz}{dy}\right)^2\right)^{\frac{3}{2}}. \tag{12}$$

The dimensionless BAE in this case can be presented as follows

$$\frac{\frac{\partial^2 \tilde{z}}{\partial y^2}}{\left(1 + \left(\frac{\partial \tilde{z}}{\partial y}\right)^2\right)^{\frac{3}{2}}} = 1 + Bo \cdot z, \tag{13}$$

where $\tilde{z} = z/r$ is the dimensionless $z$ coordinate, $Bo = \Delta \rho g r^2/\gamma = r^2/\kappa^2$ is dimensional invariant, which defines the droplet shape in gravity field, called Bond number.

In our case the parameters, which can be correlated with the results of MD simulation are the volume, the height and the contact radius of a droplet. Therefore, it is better to rewrite the Equation 12 in the coordinate system where the origin is located on the contact radius (point (0, H)) and axis z directed in opposite site than in previous case (**Figure 6**). In this case, it can be presented as follows:

$$\frac{\frac{d^2z}{dy^2}}{\left(1 + \left(\frac{dz}{dy}\right)^2\right)^{\frac{3}{2}}} = \frac{\Delta \rho g}{\gamma}\left(\frac{v}{2} - z(y)R^2\right) + \sin\theta, \tag{14}$$

Additionally, the condition of the volume conserving gives the following dependence

$$\sin\theta - \frac{R}{r} = \frac{\Delta \rho g}{\gamma}\left(H \cdot R - \frac{v}{2}\right), \tag{15}$$

The **Figure 7** presents the comparison of the droplet contact radius and radius at the droplet top evaluated with the MD simulations (points) and with Equation 14 and 15. As one can see in **Figure 6**, there are some differences between the results of the simulations and the analytical approach. Especially the contact area is systematically overestimated with the BAE model and the discrepancy increases in increasing the intensity of the external field. These differences can be overcome with the modification of volume used in the Equation 14 and 15 and Equation 2.

The dependences of the droplet height and contact radius as a function of external force intensities without (dashed-line) and with (solid-line) considered adsorption are presented in **Figure 7a** and **7b** respectively. As one can see, the solid line fits better to the MD simulations. Therefore, the adsorption layer is an important factor that should be taken into account when dealing with nanoscale water droplets.

## 4. Discussion and conclusions

In our study, we have considered a nanoscale droplet in an external homogeneous field. We prove that the macroscale approach based on Young-Laplace equation has limitations in the description of the droplet shape. In particularly, in addition to the external field, one should take into account the interatomic interaction of the atoms of the liquid with the atoms of the substrate at the interface between the two phases. A thin adsorbed layer with higher or lower density, depending the interaction between the liquid and the solid, compared to the liquid bulk value under the same conditions appears. The presence of this layer leads to the droplet contraction/expansion, which can be more or less important depending on the volume to surface ratio. In figure 8 the dependence of the ratio of the number of adsorbed particles to the total one as a function of the droplet radius is presented. In particular, for droplets with radius bigger than 100 nm, this ratio is equal to 0.1 %, and the adsorption does not significantly influence the size of the droplet (**Figure 8**).

We propose a new phenomenological model that takes into account the adsorption of liquid molecules on a solid substrate. We show also that the application of a homogenous external field allows to tune the contact area between liquid and solid for droplets with constant number of particles. As a result a reliable evaluation of the specific adsorption of the liquid atoms/molecules on a solid substrate can be evaluated.

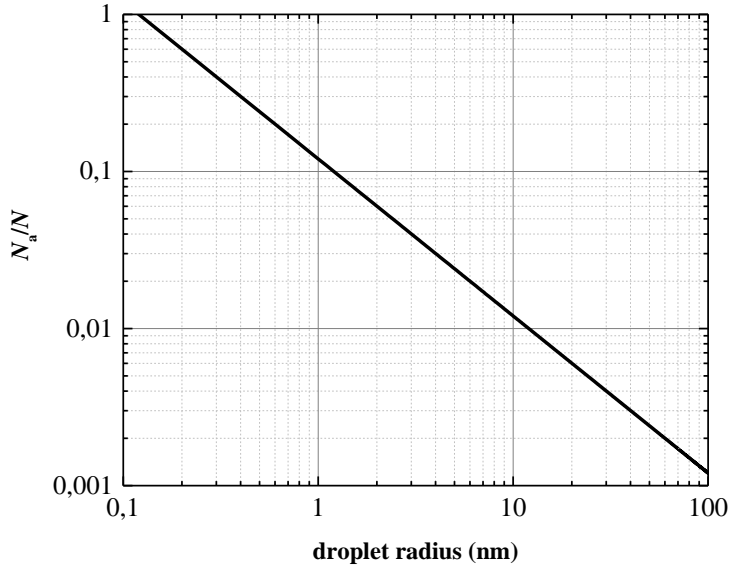

**Figure 8.** Dependence of the number of the adsorbed particles to the total one as a function of the droplet radius.

Thus, one can predict the volume contraction/expansion of the solution of a liquid with embedded nanoparticles with atomistic simulations. From the other hand, the presented approach can be easily adopted for the inverse problem of the tuning of the potential interaction between solids and liquids. For example, the parameterization of interatomic potential can be obtained by fitting of macroscale experiment of the contraction/expansion of the mentioned above solution. In this case, the Equation 11 can be considered as a dependence of the mixture volume on the specific surface of the nanoparticles.

### Acknowledgements

This work was supported by the Institute CARNOT ICEEL for the project "CAMTRASTE" (Controle et AMelioration des TRAnsferts par Structuration des surfaces d'Echange). Calculations were performed on ERMIONE cluster (IJL/LEMTA). We also thank Alain McGaughey, and Thierry Biben for fruitful discussions.